\documentclass[a4paper,12pt]{article}
\usepackage[super,compress]{cite}
\usepackage{mathrsfs}
\usepackage{amsmath}
\usepackage{amssymb}
\usepackage{graphicx}
\usepackage{bm}
\title{Behavior of light polarization in photon-scalar interaction}
\author{Azizollah Azizi and Soudabe Nasirimoghadam\\
	Physics Department, College of Sciences,\\
	 Shiraz University, Shiraz, Iran}
\begin{document}
\maketitle
\begin{abstract}
Quantum theories of gravity help us to improve our insight into the gravitational interactions. Motivated by the interesting effect of gravity on the photon trajectory, we treat a quantum recipe concluding a classical interaction of light and a massive object such as the sun. We use the linear quantum gravity to compute the classical potential of a photon interacting with a massive scalar. The leading terms have a traditional $1/r$ subordinate and demonstrate a polarization-dependent behavior. This result challenges the equivalence principle; attractive and/or repulsive interactions are admissible.
\end{abstract}
\section{\label{sec:intro} Introduction}
Regarding the direct detection of gravitational waves by the Advanced Laser Interferometer Gravitational-wave Observatory (LIGO), one finds that the Newtonian law is not suitable for the high gravity regimes such as black hole systems.\cite{LIGO1, LIGO2} In addition, one of the challenging subjects in our solar system is the investigation of Newtonian deviation of gravitational potential. In this regard two interesting theories were presented, the so-called MOND and Einstein general relativity.\cite{Milg,FaMc,Eins1916}. These theories predict deviations of 1/r law and deflection of light by the sun as well.
\par
Quantum gravity theories can also be used to obtain quantum corrections for Newtonian potential. Investigating the various interactions between photons and other fields, the gravitational potential may be derived using the field theory approach. These studies provide some modifications to the potential.
\par
In this paper, we investigate the effect of light polarization on the interaction of photons with a massive body such as the Sun. Surprisingly, we can obtain a new additional term which is originated from the light polarization. It is notable that such additional polarization term can be expressed based on the quantum prescription.
\par
A complete theory of the quantum gravity is not performed yet, however, endeavors are heavily pursued toward this end. Most candidate theories for the quantum gravity, just predict an inappreciable gravitational effect on small extents, where the standard model works very well. So, one may regard that quantum gravity is a mere theoretical scramble?
\par
Graviton---a massless boson of spin 2---is the common point of all quantum gravity models, that mediates the gravitational interactions.\cite{wien65} Between the different theories of quantum gravity, the effective field theory is capable of renormalization,\cite{don94, don95} and could be realized through experimental means. 
\par
The problem of divergences in quantum gravity, specially the infrared divergences, has been discussed by many authors in several methods.\cite{wein2,NS2011,Don-Tor,white,akhoury,Don-Hol_00,dixon1,becher,dixon2} In the case of massless particles coupled to gravity, infrared divergences---soft-infrared and collinear singularities---arise, originating from long-range low-energy degrees of freedom. Weinberg showed that the scattering of spin-zero fields is infrared finite, even the fields are massless.\cite{wein2} Moreover, he showed that the collinear divergences vanish after the sum over all diagrams with Eikonal approximation, and the remaining soft infrared divergences are removable as well. References \citen{wein2} and \citen{NS2011} have emphasized that the infrared divergences in all levels originate from one-loop divergences, and these divergences at one-loop level can be eliminated.\cite{Don-Tor} ٍEvidence shows the gravitational effective theory is well-behaved in the infrared, and there is no reason to be worried about the divergences, especially at tree-level.
\par
An effective quantum field theory of gravity is written in the weak gravitational field, where the gravity is supposed to be a linear theory. Based on the effective quantum field theory, one may drive the appropriate vertices and propagators to obtain the scattering amplitudes by using the Feynman rules. The scattering amplitudes contain two types of components, analytic and non-analytic terms. The analytic terms correspond to the local short-distance interactions, while the non-analytic terms are responsible for non-local long rang interactions described by the non-relativistic potential.\cite{don94, don94_2}
\par
In Ref.~\citen{bdh02}, up to one loop calculation and considering given non-analytic terms, the quantum corrections to the Newtonian potential between two massive scalar object are obtained. For two spin-1/2 charged massive fermions,\cite{butt06} and two massive charged scalars,\cite{faller07} scattering amplitudes are calculated for both Newtonian and Coulomb potentials. The final results of all papers show that they are in agreement upon Newtonian and post-Newtonian terms as predicted by general relativity. In Refs.~\citen{hr08_1} and \citen{hr08_2}, authors have calculated the gravitational and electromagnetic-gravitational effects of long distance potential for two massive bodies of different spins, wherein they have recognized spin-independent, spin-orbit, and spin-spin dependent contributions to potential. 
\par
In addition, authors of Ref.~\citen{pdv14} have been used some modern techniques employing spinor-helicity variables and on-shell unitarity method at the one-loop level for computing post-Newtonian and quantum corrections of the gravity. Later in Refs.~\citen{bdhp15} and \citen{bdhp16}, they applied this method to study the scattering of light and light-like particles of different spins (spin-0, spin-1/2, spin-1), with a heavy scalar (spin-0) particle such as the sun, calculating the bending angle of the particle grazing the sun. Regardless of spins, computations are in agreement with the general relativity up to the post-Newtonian corrections for all the cases. The quantum corrections come afterwards, where they depend on the spin of particles.
\par
Although the cited papers above (and many others) have done the calculations up to the one loop, in this paper we calculate the scattering amplitude only in tree level. In fact, our emphasized conclusion is resulted from tree level. One may use the loop diagrams and the required vertices to compute the loop corrections. For the sake of simplicity, we do not carry out this job here. 
\par
This paper is organized as follow: in section \ref{II}, we introduce the effective theory of quantum gravity. Section \ref{III} is devoted to explaining the rule of extracting the classical potential of gravity from scattering amplitude. In section \ref{IV}, we calculate the scattering amplitude of a photon and a massive scalar thoroughly. We obtain the potential in the classical limit and introduce the leading terms for the interaction of light and a massive body in section \ref{V}. At last, we finish the paper with some conclusions.
%
\section{Effective Field Theory of Gravity \label{II}}
We consider the quantum electrodynamics and the massive scalar field $\Phi$ are coupled to General Relativity via the Lagrangian
\begin{equation}
    \mathscr{L} = \sqrt{-g}\left[ \frac{2}{\kappa^2}\mathcal{R} - \frac{1}{4}F_{\mu\nu}F_{\rho\sigma}g^{\mu \rho}g^{\nu \sigma}
	                + \left( \frac{1}{2}g^{\mu \nu}\partial_\mu \Phi \partial_\nu \Phi-\frac{1}{2}M^2 \Phi^2\right)\right]
	                             + \cdots \label{equ1}
\end{equation}
where $g$ is the determinant of the metric tensor $g^{\mu\nu}$, $\mathcal{R}$ is the Ricci scalar, the electromagnetic field is $F_{\mu\nu}=\partial_\mu A_\nu - \partial_\nu A_\mu$ and the constant $\kappa$ is related to the gravitational constant as  $\kappa^2=32\pi G$. To linearize $ \mathscr{L}$, we suppose $g_{\mu\nu}=\bar{g}_{\mu\nu}+\kappa h_{\mu\nu}$, where $\bar{g}_{\mu\nu}$ is a background metric, and $h_{\mu\nu}$ is the fluctuation in the gravitational field. In fact, some more extra terms (gauge-fixing and ghost field Lagrangians) are needed to quantize the gravitational field.\cite{don94}
\par
In this paper, we worked in the Natural Units, so we choose $\hbar=c=1$, and the Minkowski metric is $\eta_{\mu\nu}=\textrm{diag}(+1,-1,-1,-1)$.
So one can expand the right-hand side of Eq.~(\ref{equ1}) with respect to $h_{\mu\nu}$ (or $\kappa$), and keep the required leading order terms for the linear theory. Then, it is straightforward to write down the Feynman rules,\cite{don94,bohr2002} in which we will address them in the appendix, briefly.
\section{Scattering Amplitude and Gravitational Potential} \label{III}
One can use the Feynman rules to calculate the scattering amplitude $\mathcal{M}$. As stated in Ref.~\citen{don94}, scattering amplitude in momentum space is expressed in powers of ${q}^2$. The terms proportional to positive powers of ${q}^2$, are so-called analytical terms and correspond to local interactions. These terms are related to large momentum transfer, which are not in our attention in this work. The other terms in $\mathcal{M}$, are so-called non-analytical terms which are corresponding to non-local long range interactions. In the classical limit, the potential with negative powers of $r$ comes from these terms, which are our interest in this work.
\par 
The classical potential can be constructed from quantum gravity by various methods. In Ref.~\citen{mon95} author has used the technique of Wilson loop to find an expression for the potential energy in terms of vacuum expectation value of a quantized gravitational field. In addition, there is a simple and useful prescription to define the potential energy from scattering amplitudes. Moreover, in Ref.~\citen{iwa71} Iwasaki has used the Born approximation to conclude the potential via the inverse Fourier transform of scattering amplitude
\begin{equation}\label{equ13}
	V(\textbf{r})=N\int \frac{d^3q}{(2\pi)^3} e^{-i\textbf{q}\cdot\textbf{r}}\mathcal{M}(\textbf{q}),
\end{equation}
where $N$ is a normalization factor which should be fixed.
\section{Gravitational Scattering of a Photon and a Massive Scalar Field} \label{IV}
Now we want to study the scattering of a photon by a massive scalar. In this regard, we use the Feynman diagram in tree level (Fig.~\ref{fig1}) and use the Feynman rules and propagators and vertices in the appendix. The scattering amplitude is written simply as below
\begin{figure}[t!]
	\centering
	\includegraphics[scale=1]{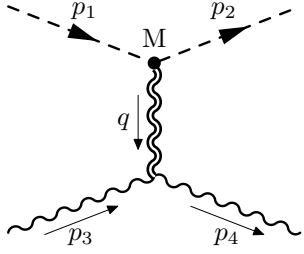}
	\caption{\label{fig1} Scalar-photon scattering at tree level.}
\end{figure}
\begin{equation}\label{equ14}
	i\mathcal{M}(q) = \epsilon^\gamma_i\ \tau_{\mu \nu \gamma \delta}(p_3,p_4)\,\epsilon^{*\delta}_f\left[ \frac{i\mathcal{P}^{\mu \nu \alpha \beta}}{q^2}\right] \tau_{\alpha \beta}(p_1,p_2, M), 
\end{equation}
where $p_1$ and $p_2$ are initial and final 4-momenta of the massive scalar field with mass $M$, and $p_3$ and $p_4$ are initial and final 4-momenta of photon
\begin{equation}\label{ener1}
	(p_1 p_1) = (p_2 p_2)=M^2, \quad (p_3 p_3) = (p_4 p_4)=0,
\end{equation}
and $\epsilon_i$ and $\epsilon_f$ are photon initial and final polarization vectors, where satisfy the relations 
\begin{equation}\label{lorc}
	(\epsilon_i\ p_3)=(\epsilon^*_f\ p_4)=0,
\end{equation}
and the transferred momentum is 
\begin{equation}\label{mom}
	q=p_1-p_2=p_4-p_3.
\end{equation}
The graviton propagator and scalar-scalar-graviton and photon-photon-graviton vertices are introduced in the appendix.
\par
Replacing from (\ref{a1})--(\ref{a3}) in Eq.~(\ref{equ14}), after contracting indices, we receive to
\begin{eqnarray}\label{equ15}
	i\mathcal{M}(q)&=&\frac{i\kappa^2}{2q^2}\left\{(p_3p_4)\left[(\epsilon_ip_1)(\epsilon^*_fp_2) +(\epsilon_ip_2)(\epsilon^*_fp_1)\right]\right.\nonumber\\
	&&+(\epsilon_i\epsilon^*_f)\left[(p_1p_3)(p_2p_4)+(p_1p_4)(p_2p_3)-(p_1p_2)(p_3p_4)\right]\nonumber\\
	&&+(p_1p_2)(\epsilon_ip_4)(\epsilon^*_fp_3)-(p_1p_3)(\epsilon^*_fp_2)(\epsilon_ip_4)-(p_1p_4)(\epsilon_ip_2)(\epsilon^*_fp_3)\nonumber\\
	&&\left.-(p_2p_3)(\epsilon^*_fp_1)(\epsilon_ip_4)-(p_2p_4)(\epsilon_ip_1)(\epsilon^*_fp_3)\right\}.
\end{eqnarray}
It is interesting to see the Eq.~(\ref{equ15}) in a more symmetrical form
\begin{eqnarray}
i\mathcal{M}(q)&=& \frac{i\kappa^2}{2q^2}\left\{ 
(\epsilon_i\epsilon^*_f)\left[(p_1p_3)(p_2p_4)+(p_1p_4)(p_2p_3)-(p_1p_2)(p_3p_4)\right]\right.\nonumber\\
&&\quad\;\; +(p_1p_2)\left[(\epsilon_ip_4)(\epsilon^*_fp_3)+(\epsilon_ip_3)(\epsilon^*_fp_4)\right]\nonumber\\
&&\quad\;\;+(p_3p_4)\left[(\epsilon_ip_1)(\epsilon^*_fp_2)+(\epsilon_ip_2)(\epsilon^*_fp_1)\right]\nonumber\\
&&\quad\;\;-(p_1p_3)\left[(\epsilon^*_fp_2)(\epsilon_ip_4) + (\epsilon^*_fp_4)(\epsilon_ip_2)\right] \nonumber\\
&&\quad\;\;-(p_1p_4)\left[(\epsilon_ip_2)(\epsilon^*_fp_3) + (\epsilon_ip_3)(\epsilon^*_fp_2)\right] \nonumber\\
&&\quad\;\;-(p_2p_3)\left[(\epsilon^*_fp_1)(\epsilon_ip_4) + (\epsilon^*_fp_4)(\epsilon_ip_1)\right] \nonumber\\
&&\quad\;\;\left.-(p_2p_4)\left[(\epsilon_ip_1)(\epsilon^*_fp_3) + (\epsilon_ip_3)(\epsilon^*_fp_1)\right] \right\}.
\end{eqnarray}
Using relations (\ref{lorc}) and (\ref{mom}), we substitute from 
\begin{eqnarray}
	(\epsilon_i p_2)&=&(\epsilon_i p_1)-(\epsilon_i q),\qquad (\epsilon_i p_4) = (\epsilon_i q),\nonumber\\
	(\epsilon^*_f p_2)&=&(\epsilon^*_f p_1)-(\epsilon^*_f q),\quad\;\; (\epsilon^*_f p_3) = -(\epsilon^*_f q), \nonumber
\end{eqnarray}
in (\ref{equ15}), and use $2(p_1q)=q^2$, and $(p_1p_4)+(p_2p_4)=(p_1p_3)+(p_2p_3)$, (after a bit rearrangement) to get
\begin{eqnarray}
i\mathcal{M}(q)&=&\frac{i\kappa^2}{2q^2} \left\{2(\epsilon_i  p_1)(\epsilon^*_f  p_1)(p_3 p_4)
-\left[(\epsilon_i  p_1)(\epsilon^*_f  q)+(\epsilon^*_f p_1)(\epsilon_i q)\right](p_3 p_4) \right.\nonumber\\
&&+\left[(\epsilon_i  p_1)(\epsilon^*_f  q)-(\epsilon^*_f p_1)(\epsilon_i q)\right]\left[(p_1 p_3)+(p_2 p_3)\right]\nonumber\\
&& +(\epsilon_i \epsilon^*_f)\left[ (p_1  p_3)(p_2  p_4)+(p_1  p_4)(p_2  p_3)-(p_1  p_2)(p_3  p_4)\right]\nonumber\\
&&\left. -(\epsilon_i  q)(\epsilon^*_f  q)\left[ (p_1  p_2)-(p_1  p_3)+(p_1  p_4)\right] \right\} \label{eqn1}.
\end{eqnarray}
\par 
We can verify the following identity
\begin{equation} \label{iden}
	\epsilon^*_{f\alpha}(\epsilon_i  q)-\epsilon_{i\alpha}(\epsilon^*_f  q)
	=\frac{2(p_3+p_4)_\alpha}{q^2}(\epsilon_i  q)(\epsilon^*_f  q)-\frac{4iE}{q^2}\,\epsilon_{\alpha \beta \gamma \delta}\,p^\beta_3q^\gamma S^\delta,
\end{equation}
where $S_\alpha$ is defined as below
\begin{equation}\label{equ20}
S_\alpha=\frac{i}{2E}\,\epsilon_{\alpha \beta \gamma \delta}\,\epsilon^{*\beta}_f\,\epsilon^\gamma_i(p_3+p_4)^\delta.
\end{equation}
Replacing from the identity (\ref{iden}) in (\ref{eqn1}) we find
\begin{eqnarray}
i\mathcal{M}(q)&=&\frac{i\kappa^2}{2q^2} \left\{ 2(\epsilon_i  p_1)(\epsilon^*_f  p_1)(p_3 p_4)
	-\left[(\epsilon_i  p_1)(\epsilon^*_f  q)+(\epsilon^*_f p_1)(\epsilon_i q)\right](p_3 p_4) \phantom{\frac{1^2}{2^2}} \right. \nonumber\\
&&\qquad +\left[-2\frac{(p_1 p_3)+(p_1 p_4)}{q^2}(\epsilon_i  q)(\epsilon^*_f  q)
+\frac{4iE}{q^2}\epsilon_{\alpha \beta \gamma \delta}p_1^\alpha p_3^\beta q^\gamma S^\delta \right] \nonumber \\
&&\qquad\quad\times \left[(p_1 p_3)+(p_2 p_3)\right]-(\epsilon_i  q)(\epsilon^*_f  q)\left[ (p_1  p_2)-(p_1  p_3)+(p_1  p_4)\right]\nonumber\\
&& \left. \;\; \phantom{\frac{1^2}{2^2}} + (\epsilon_i \epsilon^*_f)\left[ (p_1  p_3)(p_2  p_4)+(p_1  p_4)(p_2  p_3) -(p_1  p_2)(p_3  p_4)\right]\right\}\label{eqn2}.
\end{eqnarray}
Using Eqs.~(\ref{ener1}) and (\ref{mom}), and definition 
\begin{equation} \label{s}
	s=(p_1+p_3)^2,
\end{equation} 
we can simply derive the following relations
\begin{eqnarray}
	(p_1q) &=&-(p_3q)=\frac{1}{2}q^2, \nonumber\\
	(p_3p_4)&=&-\frac{1}{2}q^2, \nonumber\\
	(p_1p_4)&=&\frac{1}{2}(s-M^2+q^2), \nonumber\\
	(p_2p_4)&=&\frac{1}{2}(s-M^2), \label{rels}\\
	(p_2p_3)&=&\frac{1}{2}(s-M^2+q^2), \nonumber\\
	(p_1p_3)&=&\frac{1}{2}(s-M^2), \nonumber\\
	(p_1p_2)&=&M^2-\frac{1}{2}q^2. \nonumber
\end{eqnarray}
Replacing from Eqs.~(\ref{rels}) in Eq.~(\ref{eqn2}), after some simplifications, we achieve the following relation for the scattering amplitude
\begin{eqnarray}
\mathcal{M}(q)&=&8\pi G\left\{ -2(\epsilon_i  p_1)(\epsilon^*_f  p_1) 
	+\left[(\epsilon_i  p_1)(\epsilon^*_f  q)+(\epsilon_f^* p_1)(\epsilon_i q)\right]\phantom{\frac{1}{2}}\right.\nonumber\\
&&\qquad -\frac{4(s-M^2+\frac{1}{2}q^2)^2}{q^4}(\epsilon_i q)(\epsilon^*_f q) 
	+ \frac{8iE(s-M^2+\frac{1}{2}q^2)}{q^4}\epsilon_{\mu \beta \gamma \delta}p_1^\mu p_3^\beta q^\gamma S^\delta \nonumber\\
&& \qquad -(\epsilon_i  q)(\epsilon^*_f  q)\frac{2M^2}{q^2} \left.+(\epsilon_i \epsilon^*_f)\left[ \frac{(s-M^2)^2+sq^2}{q^2}\right]\right\}, \label{eqn3}
\end{eqnarray}
where we have replaced $\kappa^2=32\pi G$. From Eq.~(\ref{eqn3}) we can deduce the non-analytic terms of scattering amplitude 
\begin{eqnarray}
 	\mathcal{M}(q) &\simeq& -8\pi G\left\{ 
	 -(\epsilon^*_f \epsilon_i)\frac{(s-M^2)^2}{q^2} + 4(\epsilon^*_f  q)(\epsilon_i  q)\frac{(s-M^2)^2}{q^4} \phantom{\frac{(s-M^2+\frac{1}{2}q^2)}{q^4}} \right.\nonumber \\
 &&\left. 
 	+ 2(\epsilon^*_f  q)(\epsilon_i  q)\frac	{(2s-M^2)}{q^2}  -8iE \frac{(s-M^2+\frac{1}{2}q^2)}{q^4}\,\epsilon_{\mu \beta \gamma \delta}\,p_1^\mu p_3^\beta q^\gamma S^\delta \right\}. \label{nonan}
\end{eqnarray}
\par
In the following, we consider some simplifications to reach the classical limit. We assume the massive scalar is too heavy and at rest, and the collision is elastic, hence the initial and final energies of the photon remain unchanged. Under these circumstances we have
\begin{eqnarray}
	&&\mathbf{p}_1 \simeq 0 \simeq \mathbf{p}_2, \quad p_1^0 \simeq p_2^0 \simeq M,\nonumber \\
	&&p_3^0 \simeq p_4^0=E, \quad \mathbf{p}_3\cdot\mathbf{p}_3 \simeq \mathbf{p}_4\cdot\mathbf{p}_4 = E^2, \label{simp} \\
	&& \mathbf{q}=\mathbf{p}_4-\mathbf{p}_3, \quad q^0=p_1^0 - p_2^0=p_4^0 - p_3^0 \simeq 0. \nonumber
\end{eqnarray}
We suppose two transverse linear polarization for photon, so the relations (\ref{lorc}) read as
\begin{equation}\label{trans}
	\hat{\bm{\epsilon}}_i\cdot \mathbf{p}_3 = \hat{\bm{\epsilon}}_f \cdot \mathbf{p}_4 = 0. 
\end{equation}
Replacing from Eqs.~(\ref{simp}) and (\ref{trans}) in (\ref{nonan}), we are led to
\begin{eqnarray}
 \mathcal{M}(q) &\simeq& -8\pi G\left\{-\hat{\bm{\epsilon}}_f\cdot\hat{\bm{\epsilon_i}}\frac{(s-M^2)^2}{\mathbf{q}^2}
 + 4(\hat{\bm{\epsilon}}_f\cdot \mathbf{q})(\hat{\bm{\epsilon}}_i\cdot \mathbf{q})\frac{(s-M^2)^2}{\mathbf{q}^4} \phantom{\frac{(s-M^2+\frac{1}{2}q^2)}{q^4}}\right. \nonumber \\
 && \left. - 2(\hat{\bm{\epsilon}}_f\cdot \mathbf{q})(\hat{\bm{\epsilon}}_i\cdot \mathbf{q})\frac	{(2s-M^2)}{\mathbf{q}^2}
  -8iE \frac{(s-M^2 - \frac{1}{2}\mathbf{q}^2)}{\mathbf{q}^4}\,\epsilon_{\mu \beta \gamma \delta}\,p_1^\mu p_3^\beta q^\gamma S^\delta \right\}. \label{class1}
\end{eqnarray}
From Eqs.~(\ref{ener1}), (\ref{s}) and (\ref{simp}), we can investigate the following approximations
\begin{eqnarray} \label{ap2}
	&& s \simeq M^2 + 2ME, \nonumber\\
	&&S_0 \simeq \frac{i}{2E}(\mathbf{p}_3 + \mathbf{p}_4)\cdot(\hat{\bm{\epsilon}}_i\times\hat{\bm{\epsilon}}_f),\nonumber \\
	&&\mathbf{S} \simeq i\hat{\bm{\epsilon}}_i\times\hat{\bm{\epsilon}}_f, \\
	&&\epsilon_{\mu \beta \gamma \delta}\,p_1^\mu p_3^\beta q^\gamma S^\delta \simeq -M\epsilon_{ijk}p_3^i q^j S^k = M \mathbf{S}\cdot \mathbf{q}\times \mathbf{p}_3.\nonumber
\end{eqnarray}
So Eq.~(\ref{class1}) is simplified as below
\begin{eqnarray}
\mathcal{M}(q) &\simeq& -8\pi G\left\{-\hat{\bm{\epsilon}}_f\cdot\hat{\bm{\epsilon_i}}\frac{4M^2 E^2}{\mathbf{q}^2}\right.
 + 4(\hat{\bm{\epsilon}}_f\cdot \mathbf{q})(\hat{\bm{\epsilon}}_i\cdot \mathbf{q})\frac{4M^2 E^2}{\mathbf{q}^4} \nonumber \\
&& -2(\hat{\bm{\epsilon}}_f\cdot \mathbf{q})(\hat{\bm{\epsilon}}_i\cdot \mathbf{q})\frac	{(M^2 + 4ME)}{\mathbf{q}^2}
\left. - 4iEM \frac{(4ME-\mathbf{q}^2)}{\mathbf{q}^4}\, \mathbf{q}\times \mathbf{p}_3 \cdot \mathbf{S} \right\}. \label{class2}
\end{eqnarray}
Taking into account Eq.~\ref{class2}, it is notable that the first term is the Newtonian potential term and the second and fourth terms come from identity (\ref{iden}) and former has a part like the Newtonian term, while the third and fourth terms related to the angular momentum. 
\section{Potential}\label{V}
Now we are in a position to calculate the classical gravitational potential. Inserting Eq.~(\ref{class2}) into Eq.~(\ref{equ13}) and using the Fourier transformation (see Eqs.~(\ref{a4})--(\ref{a8}) in the appendix for more details), we obtain 
\begin{eqnarray}
V(r) &\simeq& -8\pi G N\left\{-\hat{\bm{\epsilon}}_f\cdot\hat{\bm{\epsilon_i}}\frac{4M^2 E^2}{4\pi r}\right. \nonumber \\
&& \qquad\qquad\;\, + 4(\hat{\bm{\epsilon}}_f\cdot \hat{\bm{\epsilon}}_i)\frac{4M^2 E^2}{8\pi r} 
			- 4(\hat{\bm{\epsilon}}_f\cdot \mathbf{r})(\hat{\bm{\epsilon}}_i\cdot \mathbf{r})\frac{4M^2 E^2}{8\pi r^3} \nonumber \\
&& \qquad\qquad\;\, - 2\frac{1}{4\pi}(\hat{\bm{\epsilon}}_f\cdot \hat{\bm{\epsilon}}_i)\frac{(M^2 + 4ME)}{r^3}
	+ 2\frac{3}{4\pi}(\hat{\bm{\epsilon}}_f\cdot \mathbf{r})(\hat{\bm{\epsilon}}_i\cdot \mathbf{r})\frac{(M^2 + 4ME)}{r^5} \nonumber\\
&& \qquad\qquad\;\, -4iEM \frac{-i}{8\pi}\frac{4ME}{r}\, \mathbf{r}\times \mathbf{p}_3 \cdot \mathbf{S}
	\left. +4iEM \frac{-i}{4\pi}\frac{1}{r^3}\, \mathbf{r}\times \mathbf{p}_3 \cdot \mathbf{S} \right\}. \label{pot1}
\end{eqnarray}
In our case study, the normalization factor for the massive scalar and massless photon is chosen as $N=1/\surd(2E_12E_2 2E_3 2E_4)=1/4ME$. Simplifying Eq.~(\ref{pot1}), we find
\begin{eqnarray}
 V(r) &\simeq& \hat{\bm{\epsilon}}_f\cdot\hat{\bm{\epsilon_i}}\frac{-2G M E}{r}
	+ (\hat{\bm{\epsilon}}_f\cdot \mathbf{r})(\hat{\bm{\epsilon}}_i\cdot \mathbf{r})\frac{4G M E}{r^3} \nonumber \\
&& + (\hat{\bm{\epsilon}}_f\cdot \hat{\bm{\epsilon}}_i)\frac{G(M/E + 4)}{r^3} 
	 - (\hat{\bm{\epsilon}}_f\cdot \mathbf{r})(\hat{\bm{\epsilon}}_i\cdot \mathbf{r})\frac{3G(M/E + 4)}{r^5} \nonumber\\
&&  + \frac{4GM E}{r}\, \mathbf{r}\times \mathbf{p}_3 \cdot \mathbf{S}
	 - \frac{2G}{r^3}\, \mathbf{r}\times \mathbf{p}_3 \cdot \mathbf{S}. \label{pot2}
\end{eqnarray}
Eq.~(\ref{pot2}) is supposed to be valid in the classical limit, since we have used the linear approximation of gravity. These expressions show that, interestingly, the polarization has considerable effects on the gravitational potential. Comparing our result with Ref.~\citen{hr08_1} (for a massive spin-1 particle in the field of a massive scalar particle), we find that the first term in Eq.~(\ref{pot2}) is in agreement with the general relativity. The second term is the main interesting result of this paper. This is a $1/r$ term, that will not be affected by a further consideration on calculating the loop corrections. Against the first term, this is a repulsive term of the same order. In Ref.~\citen{hr08_1}, within the non-relativistic limits, the transferred momentum is neglected in comparison with the masses in theory, so the results lack such a term.
\par
In Refs.~\citen{bdhp15} and \citen{bdhp16} (see Ref.~\citen{bai17} as well), for massless particles interacting massive particles, the computations are done independently of spin and polarization orientations. For the bending angle of the light grazing the sun, the leading and even the next to the leading terms are in full agreement with the general relativity.
\section{Conclusions \label{VI}}
In quantum field theory, the interaction of figure \ref{fig1} is computed for a time duration of $\Delta t =\hbar/E$. So, in the classical limit, we can accept that the polarization of the photon does not change within such a $\Delta t \sim 0$, i.e.~we can choose $\hat{\bm{\epsilon}}_i=\hat{\bm{\epsilon}}_f\equiv \hat{\bm{\epsilon}}$. In this limit $\mathbf{S}$ (Eqs.~(\ref{ap2})) vanishes, therefore the last two terms in Eq.~(\ref{pot2}) are annihilated. For the large $r$, the first two terms in Eq.~(\ref{pot2}) are the leading terms in negative powers of $r$,
\begin{equation}\label{oz}
	V(r) \simeq -\frac{2GME}{r} + (\hat{\bm{\epsilon}}\cdot \mathbf{r})^2\frac{4GME}{r^3}.
\end{equation}
That is a distinguished result. We emphasize that Eq.~(\ref{oz}) is derived from tree-level amplitude, and loop corrections do not change the result. The first term in Eq.~(\ref{oz}) is the usual inverse distance law of gravitational potential for a photon in the field of a mass $M$ predicted by general relativity. The second term in Eq.~(\ref{oz}) is a polarization dependent expression of $1/r$ dependence.
\par
The first thing that comes into sight is the difference between the signs of two terms; while the first term is attractive, the second term is repulsive. Meanwhile, the coefficient of the new term is twice as large as the first. Eq.~(\ref{oz}) can be rewritten in the form 
\begin{equation} \label{oz1}
	V(r) \simeq -\frac{2GME}{r}(1 - 2\cos^2{\gamma}) = \frac{2GME}{r}\cos(2\gamma),
\end{equation}
where $\gamma\in[0,\pi]$ is the angle between $\mathbf{r}$, the position vector of the photon (light), and its polarization. Depending on the orientation of the polarization vector with the line connecting the photon and the scalar particle, the net force on the photon can be either attractive or repulsive.
\par
According to the potential (\ref{oz}), we may assume a plane of motion for a photon around the ``sun''. With the polarization of photon perpendicular to this plane at far past, it remains perpendicular to the plane at a later time. In this case, we dealt with the usual attractive gravitational force. Accordingly, the bending of light around the sun is calculated via the first term in (\ref{oz}), which is predicted by the general relativity a century ago. 
\par
If the polarization vector of light stays in the plane of motion, the second term plays an important role. To illustrate the problem, suppose a beam of light is coming far from the sun, where the position vector of light and its momentum vector are parallel. So the polarization vector of light is perpendicular to the position vector, and the light feels only an attractive force from the sun. With light becoming closer, the polarization vector makes an angle $\theta$ with the position vector $\mathbf{r}$, so a contradictory force presents itself, playing a role in deflection. When the light has the smallest distance from the sun, i.e.~at perihelion that $\mathbf{r}$ is perpendicular to the light path, the net force is repulsive, and its magnitude is equal to the traditional attractive force.
\par
Potential (\ref{oz}), makes different patterns for deflection of light. These new patterns can be tested in a solar eclipse. Potential (\ref{oz}) affects the gravitational lensing as well. We will discuss these items in a next paper.
\appendix
\section{Propagators, Vertices and useful Fourier transformations \label{app1}}
\subsection{Graviton propagator}
\begin{eqnarray}
	\parbox{60pt}{\includegraphics[scale=.75]{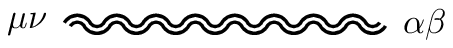}} \qquad\qquad &=& \frac{i\mathcal{P}^{\mu \nu \alpha \beta}}{q^2},\nonumber\\
	\mathcal{P}^{\mu \nu \alpha \beta} &\equiv& \frac{1}{2}(\eta^{\mu \alpha}\eta^{\nu \beta}+\eta^{\mu \beta}\eta^{\nu \alpha}-\eta^{\mu \nu}\eta^{\alpha \beta}).\; \label{a1}
\end{eqnarray}
\subsection{Scalar-scalar-graviton vertex}
\begin{eqnarray}
	\parbox{60pt}{\includegraphics[scale=.75]{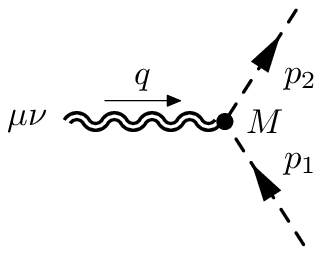}} \qquad &=& \, \tau_{\mu \nu}(p_1,p_2,M)\nonumber\\
	&=&-\frac{i\kappa}{2}\left[ p_{1\mu}p_{2\nu}+p_{1\nu}p_{2\mu}-\eta_{\mu \nu}(p_1\cdot p_2-M^2)\right].\; \label{a2}
\end{eqnarray}

\subsection{Photon-photon-graviton vertex}
\begin{eqnarray}
	\parbox{60pt}{\includegraphics[scale=.75]{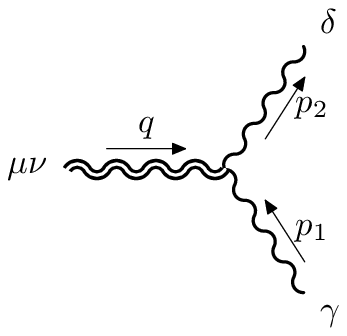}} \qquad &=& \, \tau_{\mu \nu \gamma \delta}(p_1,p_2)\nonumber\\
    &=& i\kappa\left \{\mathcal{P}_{\mu \nu \gamma \delta}(p_1\cdot p_2)+\frac{1}{2}\left[\eta_{\mu \nu}p_{1\delta}p_{2\gamma}+\eta_{\gamma \delta}(p_{1\mu}p_{2\nu}+p_{1\nu}p_{2\mu})\right.\right.\nonumber\\
    &&\left.\left.-(\eta_{\mu \delta}p_{2\gamma}p_{1\nu}+\eta_{\nu \delta}p_{2\gamma}p_{1\mu}+\eta_{\nu \gamma}p_{2\mu}p_{1\delta}+\eta_{\mu       	\gamma}p_{2\nu}p_{1\delta})\right]\phantom{\frac{1}{2}} \!\!\!\!\!\right\}.\; \label{a3}
\end{eqnarray}
\subsection{Useful Fourier Transformations}
\begin{eqnarray}
	\int \frac{d^3q}{(2\pi)^3}\,e^{-i\mathbf{q}\cdot \mathbf{r}} \left(\frac{1}{\mathbf{q}^2}\right)&=&\frac{1}{4\pi r}, \label{a4} \\
	\int \frac{d^3q}{(2\pi)^3}\,e^{-i\mathbf{q}\cdot \mathbf{r}} \left(\frac{q_i}{\mathbf{q}^2}\right)&=&-\frac{ir_i}{4\pi r^3}, \label{a5}\\
	\int \frac{d^3q}{(2\pi)^3}\,e^{-i\mathbf{q}\cdot \mathbf{r}} \left(\frac{q_iq_j}{\mathbf{q}^2}\right)&=& -\frac{1}{4\pi}\left(3\frac{r_ir_j}{r^5}-\frac{\delta_{ij}}{r^3}\right), \label{a6}\\
	\int \frac{d^3q}{(2\pi)^3}\,e^{-i\mathbf{q}\cdot \mathbf{r}} \left(\frac{q_iq_j}{\mathbf{q}^4}\right)&=&\frac{1}{8\pi}\left(\frac{\delta_{ij}}{r}-\frac{r_ir_j}{r^3}\right), \label{a7} \\
	\int \frac{d^3q}{(2\pi)^3}\,e^{-i\mathbf{q}\cdot \mathbf{r}} \left(\frac{q_i}{\mathbf{q}^4}\right)&=&-\frac{ir_i}{8\pi r}. \label{a8}
\end{eqnarray}
\section*{Acknowledgments}
The authors would like to thank John F.~Donoghue and Barry R.~Holstein for kindly answering our questions.

\end{document}